\documentclass[runningheads]{llncs}
\usepackage{graphicx}
\usepackage{cite}
\usepackage{hyperref}
\usepackage{amsmath,amssymb,amsfonts}
\usepackage{algorithmic}
\usepackage{graphicx}
\usepackage{subfig}
\usepackage{textcomp}
\usepackage{booktabs}
\usepackage{footnote}
\usepackage{fancyhdr}
\usepackage[para]{threeparttable}
\makesavenoteenv{tabular}
\makesavenoteenv{table}
\usepackage[table]{xcolor}
\usepackage{multirow}
\begin{document}
\title{Inter and Intra Signal Variance in Feature Extraction and Classification of Affective State\thanks{This publication has emanated from research supported in part by a Grant from Science Foundation Ireland under Grant number 18/CRT/6222}}

%
%
\author{Zachary Dair\inst{1}\orcidID{0000-0003-0203-0396} \and
Samantha Dockray\inst{2}\orcidID{0000-0002-0804-8362} \and
Ruairi O'Reilly\inst{1}\orcidID{0000-0001-7990-3461}}
\authorrunning{Z. Dair et al.}
%
\institute{1. Munster Technological University, Cork, Ireland\\
2. University College Cork, Cork, Ireland\\
\email{zachary.dair@mycit.ie}\\
\email{s.dockray@ucc.ie}\\
\email{ruairi.oreilly@mtu.ie}}
\maketitle              
\begin{abstract}
Psychophysiology investigates the causal relationship of physiological changes resulting from psychological states. There are significant challenges with machine learning-based momentary assessments of physiology due to varying data collection methods, physiological differences, data availability and the requirement for expertly annotated data. Advances in wearable technology have significantly increased the scale, sensitivity and accuracy of devices for recording physiological signals, enabling large-scale unobtrusive physiological data gathering. This work contributes an empirical evaluation of signal variances acquired from wearables and their associated impact on the classification of affective states by (i) assessing differences occurring in features representative of affective states extracted from electrocardiograms and photoplethysmography, (ii) investigating the disparity in feature importance between signals to determine signal-specific features, and (iii) investigating the disparity in feature importance between affective states to determine affect-specific features. Results demonstrate that the degree of feature variance between ECG and PPG in a dataset is reflected in the classification performance of that dataset. Additionally, beats-per-minute, inter-beat-interval and breathing rate are identified as common best-performing features across both signals. Finally feature variance per-affective state identifies hard-to-distinguish affective states requiring one-versus-rest or additional features to enable accurate classification.


\end{abstract}
\section{Introduction}
\label{sec:introduction}
A significant goal of Affective Computing is to improve human-to-computer interaction by providing a system with a level of emotional intelligence that aids natural communications and is capable of including emotional components \cite{PICARD200355}. This has commonly been approached by deriving emotional states from speech, facial expressions, gestures and body posture analysis. However, utilising physiological signals to communicate psychological information is a recent exploration in the domain, likely due to the increased accessibility of signals from wearables.

 A physiological signal represents an individual's biological processes derived from core aspects of human biology. These signals can enable diagnostics, for instance, analysing heart rate (HR) to detect arrhythmia \cite{LUZ2016144}. Psychological analysis can also be enabled as mental states originating from unconscious effort typically present a noticeable physiological change in the relevant human system \cite{4441720}. The combined analysis enables a richer understanding of individuals in terms of their mental and physical health \cite{Dzedzickis_2020}.
 
Psychological states are complex processes comprised of several components, including feelings, cognitive reactions, behaviour and thoughts\cite{Agrafioti_2012}. Mapping psychological states to individual experience provides valuable information regarding well-being, health (physical and mental), social contexts, experiences and emotions \cite{dockray_psycho_bio_correlates}.

Electrocardiograms (ECG) are physiological signals that measure the electrical activity of the heart. Typically recorded in a clinical setting using multiple electrodes attached to the individual. Photoplethysmography (PPG) is a physiological signal used to measure heart activity through variations in the blood volume of the skin, using a light-emitting-diode and photodetector. Wearable devices predominately utilise PPG to monitor heart activity. However, recently advanced wearables have included ECG capabilities for a limited number of commercial off-the-shelf (COTS) devices.

Data variances occur when recording ECG and PPG due to differing sensor placement and signal granularity \cite{7319746, Dzedzickis_2020}. A lower sampling frequency is commonly used in PPG compared to ECG to reduce battery consumption in COTS devices. Such variances are under-recognised in the field of psychophysiology.

This work investigates the impact of signal variances occurring in ECG and PPG signals acquired from wearable devices for classifying affective states by addressing the following research aims: (i) To assess differences in features representative of affective states on a per signal basis, (ii) To investigate the disparity in precedence ordering of feature importance per signal, and (iii) To investigate the disparity precedence ordering of feature importance per affective state.

These aims inform the development of machine learning (ML) pipelines for classifying affective states. Utilising feature variance per signal to identify abnormal signal activity or similar affective states which are causing reduced classification accuracy. In conjunction, feature importance is utilised to provide insights into feature selection, aiding performance in tailored signal- or affect-specific approaches.

\section{Related Work}
\label{sec:relatedWork}

\subsection{Heart-Related Physiological Signals}
The prevalence of heart-related data in wearable devices stems from a desire to monitor health through arrhythmia detection and HR as a measure of fitness \cite{LUZ2016144}. As the heart is controlled involuntarily through the autonomic nervous system (ANS), it facilitates identifying relationships between involuntary physiological changes in heart activity and psychological states such as emotions or behaviour. Multiple psychophysiological theories aim to explain this relationship, such as Polyvagal Theory \cite{Porges1994-oy}, which proposes that the ANS provides the neurophysiological substrates for adaptive behavioural strategies \cite{Porges2009}.

Heart activity is complex to capture. In medicine, the gold standard utilises a 12-lead ECG, resulting in comprehensive data recorded from multiple electrodes on the human body. However, in ambulatory research and daily life, this method is not feasible. Typically research-grade (RG) equipment uses several electrodes, commonly 3-lead ECG, and occasionally includes PPG as an additional measure. COTS devices tend to rely solely on PPG to monitor heart activity. However, with recent advances, top-of-the-range smart-watches (Apple Watch 4-9, Galaxy Active 2, Fitbit) include a 1-lead ECG, which is promising for portable ECG analysis \cite{SAGHIR202030}.

Additional physiological signals such as electrodermal activity (EDA), respiration, skin temperature, electromyogram (EMG), and electrooculogram (EOG) have demonstrated potential for affective state detection \cite{Dzedzickis_2020,Shu2018}; however, due to additional sensor requirements they are excluded from this work.

Numerous studies of affective states conduct custom data collection, providing precise control over the psychological domain explored. Varied stimuli have aided the elicitation of psychological states, for example, images, movie clips, music, and dedicated tasks to elucidate stress, such as the Trier Social Stress Test \cite{WESAD_dataset,Birkett2011-wz}. As denoted in Table \ref{tab:datasets}, several open-access or on-request datasets containing ECG and PPG are available. The distinct lack of emotionally labelled ECG signals from COTS devices is likely due to the recent inclusion of ECG monitoring capabilities \cite{SAGHIR202030}.

\begin{table}[ht]
    \begin{center}
        \caption{Datasets containing affectively labelled ECG, PPG or both}
        \label{tab:datasets}
        \begin{tabular}{ccccc}
        \toprule
            Dataset & ECG & PPG & Participants & Label\\
            \midrule
            CASE\cite{CASE_dataset} & \checkmark (1000Hz) & \checkmark (1000Hz) & 30 & Aro./Val.\\
            WESAD\cite{WESAD_dataset} & \checkmark (700Hz) & \checkmark (64Hz) & 15 & B, S, A, M\\
            DREAMER\cite{Dreamer_dataset} & \checkmark (256Hz) & x & 23 & Aro./Val./Dom.\\
            SWELL\cite{koldijk_swell_2014} & \checkmark (2048Hz) & x & 25 & S, Aro./Val./Dom.\\
            DEAP\cite{DEAP_dataset} & x & \checkmark (256Hz) & 32 & Aro./Val.\\
            \midrule
            \multicolumn{5}{c}{B: Baseline, 
                    S: Stress,
                    A: Amusement,
                    M: Meditation
                    }\\
            \multicolumn{5}{c}{
            Aro: Arousal,
            Val: Valence,
            Dom: Dominance
            }\\
        \bottomrule
        \end{tabular}
    \end{center}        
\end{table}

\begin{table}[h]
    \centering
    \caption{Handcrafted, automated and statistical features utilised for affective state classification. Note the divergence between features used by PPG and ECG.}
    \label{tab:overview_features}
    \resizebox{\textwidth}{!}{
    \begin{tabular}{lcccccccc|ccccccc|cc}
    \toprule
    &\multicolumn{8}{c|}{ECG}&\multicolumn{7}{c|}{PPG}&\multicolumn{2}{c}{Combined}\\
         & \cite{sarkar_self-supervised_2021} & \cite{Agrafioti_2012} & \cite{cai_jing} & \cite{burcu_cinaz} & \cite{nardelli_recognizing_2015} & \cite{hsu_automatic_2020} & \cite{Dissanayake_2019} & \cite{SAYEDISMAIL20223539} &
         
         \cite{Goshvarpour2020} & \cite{mukherjee_real-time_2022} & \cite{GOSHVARPOUR2018} & \cite{prerita_kalra} & \cite{tucker_catching_2021} & \cite{2019_seop_lee} & \cite{SAYEDISMAIL20223539} &
         
         \cite{CHEEMA2019493} & \cite{filippini_chiara} \\
         
        \midrule
        \multicolumn{9}{l|}{HRV Time:}&\multicolumn{7}{l|}{}\\
        \hspace*{0.2cm} HR & x & x & x & \checkmark & x & \checkmark & x & \checkmark &
        x & x & x & \checkmark & x & x & \checkmark &
        x & x \\
        
        \hspace*{0.2cm} IBI & x & x & x & x & x & x & x & \checkmark &
        x & x & x & x & x & x & \checkmark &
        \checkmark & x \\
        
        \hspace*{0.2cm} RR & x & x & x & \checkmark & \checkmark & \checkmark & \checkmark & \checkmark &
        x & x & x & x & x & x & \checkmark &
        x & \checkmark \\
        
        \hspace*{0.2cm} SD & x & x & \checkmark & \checkmark & \checkmark & \checkmark & \checkmark & \checkmark &
        x & x & x & x & x & x & \checkmark &
        x & \checkmark \\
        
        \hspace*{0.2cm} P-QRS-T & x & x & \checkmark & x & x & x & \checkmark & \checkmark &
        x & x & x & x & x & x & \checkmark &
        x & x \\
        
        \midrule
        \multicolumn{9}{l|}{HRV Frequency:}&\multicolumn{7}{l|}{}  \\
        \hspace*{0.2cm} Low Freq. & x & x & \checkmark & \checkmark & \checkmark & \checkmark & \checkmark & \checkmark &
        x & x & x & \checkmark & x & x & \checkmark &
        x & \checkmark \\
        
        \hspace*{0.2cm} High Freq. & x & x & x & \checkmark & \checkmark & \checkmark & \checkmark & \checkmark &
        x & x & x & \checkmark & x & x & \checkmark &
        x & \checkmark \\
        
        \hspace*{0.2cm} Freq. Ratios & x & x & x & \checkmark & \checkmark & \checkmark & \checkmark & \checkmark &
        x & x & x & \checkmark & x & x & \checkmark &
        \checkmark & \checkmark \\
        
        \midrule
        \multicolumn{9}{l|}{Non-Linear:}&\multicolumn{7}{l|}{}  \\
        \hspace*{0.2cm} Poin. & x & x & x & x & \checkmark & \checkmark & \checkmark & x &
        x & x & \checkmark & x & x & x & x &
        x & x \\
        
        \hspace*{0.2cm} Entr. & x & x & x & x & \checkmark & \checkmark & x & \checkmark &
        \checkmark & x & x & x & x & x & \checkmark &
        \checkmark & \checkmark \\
        
        \midrule
        \multicolumn{9}{l|}{Deep Learning:}&\multicolumn{7}{l|}{}  \\
        \hspace*{0.2cm} Model & \checkmark & x & x & x & x & x & x & x &
        x & x & x & x & \checkmark & \checkmark & x &
        x & x \\
        
        \hspace*{0.2cm} AE & x & x & x & x & x & x & x & x &
        x & \checkmark & x & x & x & x & x &
        x & x \\
        
        \midrule
        \multicolumn{9}{l|}{Additional Features:}&\multicolumn{7}{l|}{}  \\
        \hspace*{0.2cm} BR & x & x & x & x & x & \checkmark & x & x &
        x & x & x & x & x & x & x &
        x & \checkmark \\
        
        \hspace*{0.2cm} Sig. Amp. & x & x & x & x & x & x & x & \checkmark &
        x & x & x & x & x & x & \checkmark &
        x & \checkmark \\
        
        \hspace*{0.2cm} EMD & x & \checkmark & x & x & x & x & \checkmark & x &
        x & x & x & x & x & x & x &
        \checkmark & x \\
        
        \midrule
        Classifier & CNN & LDA & KNN & LDA & QDC & SVM$^b$ & Ens. & SVM &
        PNN & SVM & SVM & DNN & CNN & CNN & SVM &
        SVM$^b$ & FNN \\
        
        No. Classes & 4 & 2 & 2 & 3 & 4 & 2 & 4 & 2 &
        14$^a$ & 4$^a$ & 2 & 5 & 2 & 2 & 2 &
        2 & 4 \\
        
        Accuracy & 95\% & 89\% & 85\% & 85\% & 84\% & 82\% & 80\% & 69\% &
        100\% & 99\% & 96\% & 91\% & 83\% & 76\% & 65\% &
        93\% & 70\%\\
        
        Datasets & \cite{WESAD_dataset} & Priv. & Priv. & Priv. & Priv. & Priv. & Priv. & Priv. &
        \cite{DEAP_dataset} & \cite{WESAD_dataset} & \cite{DEAP_dataset} &  Priv. & \cite{WESAD_dataset} & \cite{DEAP_dataset} & Priv. &
        Priv. & Priv.\\
        
        \midrule
       
        \multicolumn{18}{l}{
            HRV: Heart Rate Variability,
            IBI: Inter-beat Interval,
            RR: R-R Intervals,
            SD: Successive Differences,
            }\\
        \multicolumn{18}{l}{
            Freq: Frequency,
            Poin: Poincare,
            Entr: Entropy,
            AE: Auto-Encoder,
            BR: Breathing Rate,
            Ens: Ensemble,
            }\\
        \multicolumn{18}{l}{
            Sig. Amp: Signal Amplitude,
            EMD: Empirical Mode Decomposition,
            LDA: Linear Discriminant Analysis,
            }\\
        \multicolumn{18}{l}{
            (C/D/F)NN: Convolutional/Deep/Feed-Forward Neural Network,
            QDC: Quadratic Discriminant Classifier,
            }\\
        \multicolumn{18}{l}{
            SVM: Support Vector Machine,
            KNN: K-Nearest Neighbours,
            $^a$ One Vs Rest,
            $^b$ Least-Squares SVM
            }\\
        
    \bottomrule
    \end{tabular}
    }
\end{table}

\subsection{Affective ECG Analysis}
ECG signals contain noise introduced by motion artefacts, biological differences and sensor de-attachment. Signal processing techniques such as Butterworth Bandpass, Notch filters and Empirical Mode Decomposition (EMD) are utilised to reduce the signal noise levels \cite{Agrafioti_2012}. Subsequently, features suitable for affective state classification can be extracted from the pre-processed signals.

An overview of features derivable from ECG and PPG is denoted in Table \ref{tab:overview_features}, grouped by extraction method. 
Performant ECG-based approaches typically utilise handcrafted features, particularly time-based HRV features, such as R-R intervals (RR) which are the intervals between heartbeats, successive differences (SD) and frequency-based features, such as relative, peak and absolute power of various frequency bands. Automated feature extraction is less frequently adopted, with only three of the reviewed approaches utilising deep learning or signal-processing feature extraction methods.

Recent approaches have favoured deep learning methodologies \cite{sarkar_self-supervised_2021}, achieving significant accuracies on multi-class classifications. However, older studies focusing on linear and quadratic discriminant analysis (LDA, QDA) \cite{Agrafioti_2012,nardelli_recognizing_2015, burcu_cinaz} and support vector machines (SVM) \cite{hsu_automatic_2020} remain highly relevant, achieving high accuracy for their respective classifications. Combinations of ML classifiers forming ensembles have demonstrated potential for binary classifications in emotion detection \cite{Dissanayake_2019}. In comparison to other studies, \cite{sarkar_self-supervised_2021} achieved the highest accuracy for multiple emotion detection from ECG data utilising a CNN and reported setting the new state of the art for ECG emotion detection. Despite the high performance of deep learning approaches in the literature, this work focuses on classifiers using handcrafted features.

\subsection{Affective PPG Analysis}
PPG analysis provided by COTS devices has typically focused on tracking medical conditions, physical activity, and stress. The detrimental effects of stress on human health are a significant motivator for physiological analysis and preventative healthcare research \cite{can_continuous_2019}. However, instances of PPG have demonstrated similar noise levels to ECG, with the addition of skin tone and environmental light effects impacting signal quality, requiring signal cleaning techniques.

There is no consensus on the most frequently used features from the reviewed PPG-based approaches, see Table \ref{tab:overview_features}. The most performant approach \cite{Goshvarpour2020} leverages handcrafted non-linear entropy features, followed by \cite{mukherjee_real-time_2022} using an autoencoder method for automatic feature extraction. Importantly, both handcrafted and automatically extracted features aid in achieving a high classification accuracy above 90\% \cite{Goshvarpour2020,mukherjee_real-time_2022,GOSHVARPOUR2018,prerita_kalra}.

These affective state classifications are conducted by variations of neural networks \cite{Goshvarpour2020,prerita_kalra,tucker_catching_2021,2019_seop_lee} and SVMs \cite{mukherjee_real-time_2022,GOSHVARPOUR2018,SAYEDISMAIL20223539}, which demonstrates great potential for both binary and multi-class affective state detection using PPG solely. Notably, these approaches leverage extensive signal processing to reduce signal noise and contribute to the high performances achieved.

\section{Methodology}
\label{sec:methodology}

The proposed methodology provides an approach for investigating ECG and PPG variances and the subsequent impact on affective state classification. The baseline performance of affective state classification is achieved using multiple ML classifiers per signal. The inter-signal performance variances are investigated by analysing the disparity in features between temporally aligned ECG and PPG, where the degree of feature variance is an indicator of signal quality. Inter-affective state feature variance is analysed using statistical measures to provide insights into the distribution and similarity of affective states. Feature importance is employed to identify commonalities among the best-performing features across both signals and evaluate each feature's utility for affect-specific approaches. Finally, a one-versus-rest (OVR) classification is adopted to improve performance when classifying similar affective states.

\subsection{Datasets}
For the purposes of this work, the focus was narrowed to RG physiological signals due to a lack of publicly available data for COTS devices. ``The Dataset of Continuous Affect Annotations and Physical Signals for Emotion Analysis'' (CASE)\cite{CASE_dataset} and ``The Wearable Stress and Affect Detection Dataset'' (WESAD)\cite{WESAD_dataset}, see Table \ref{tab:datasets}, were utilised in this work. The datasets were selected due to their inclusion of temporally aligned ECG and PPG with psychological annotations. Additionally, these signals were recorded using RG devices in a laboratory environment. CASE incorporates Arousal and Valence annotations, achieved by collecting joystick movement resulting from emotionally stimulating video clips. WESAD focuses on stress detection with limited affective states: a baseline state elicited from ``neutral reading'', amusement caused by comedic video clips, a Trier Social Stress Test\cite{Birkett2011-wz} to provoke stress, and a meditation stage aimed at ``de-exciting'' the individual following the amusement and stress stages.

\subsection{Pre-Processing}
ECG and PPG signals recorded per subject within these datasets span the duration of the experiment resulting in approx 91/40 minutes for WESAD/CASE. Each signal is pre-processed into 10-second windows to facilitate analysis, accomplished using a sliding window technique with a 1-second overlap. A 10-second duration was selected due to efficient performance demonstrated in \cite{sarkar_self-supervised_2021}; additionally, this duration enables low latency as classification occurs every 10-seconds and contains adequate data for feature computation.

A Butterworth-Bandpass filter is used to reduce signal noise, facilitating the extraction of selected features while maintaining a degree of ``rawness'' in the signal. This filter was adopted as it is frequently adopted in the literature and more closely aligns with COTS devices and their reduced computational power.

Once filtered and windowed, the data is aligned with the psychological annotations. For WESAD, annotations were numeric values sampled at 700Hz. Each value from 0-4 is associated with the psychological states: Transient, Baseline, Stress, Amusement and Meditation. Annotations 5-7 and Transient data are omitted as per the author's instructions \cite{WESAD_dataset}. Certain windows may include multiple emotive annotations; hence to identify the most pertinent emotion, the mean of all annotation values per window is calculated and rounded to the nearest annotation (1-4) using Euclidean distance. Alternative approaches\cite{Dissanayake_2019} omit these windows and the neighbouring segments to prevent overlap.

A similar procedure is required for CASE; the raw annotation data is provided as values on an x and y-axis representing Arousal and Valence\cite{CASE_dataset}, these values are normalised to a range of 0.5 to 9.5, and subsequently converted to discrete representations, resulting in low (0.5-5) and high (5.01-9.5) Arousal and Valence for each window.

Both signals provide capabilities to derive a wide array of handcrafted features useful for identifying affective information. This work utilises a python toolkit HeartPy \cite{VANGENT2019368} to enable extraction of HRV features from each window of data, summarised in accompanying table of Figure \ref{fig:feature_diff_WESAD}.

\subsection{HRV Feature Variance}
The feature variance approach proposed is to statistically evaluate any disparity occurring in derived features from ECG and PPG under multiple conditions. Inter-signal variance is evaluated by computing the absolute difference between an ECG-derived feature and its PPG counterpart from temporally aligned signals. This is assessed using the same window of heart activity and provides a granular analysis to aid in identifying noisy, erratic or abnormal signal activity, causing unreliable computations of features. This variance is depicted by a significant absolute difference of a feature between the two signals.

Analysing the inter-affective state variance in features enables the identification of the degree of change between states, as investigated in \cite{GOSHVARPOUR2018,hsu_automatic_2020}. The proposed methodology computes the minimum, maximum, mean and standard deviation of each feature value per affective state. Additionally outliers are identified, these are observations found in the upper and lower quartiles. This method identifies states which are complex to distinguish due to a similar feature distribution, such as meditation and relaxation. An OVR approach is adopted to convert a multi-class problem into multiple binary classifications. Using OVR, a classifier aims to identify an affective state individually from the remaining states, which increases the degree of distinction between classes.

\subsection{HRV Feature Importance}
This work adopts a game theory approach for feature importance known as ``Shapley Additive exPlanations``\cite{molnar2022}. This method computes SHAP values representing the degree of change on the classifier output caused by each individual feature, the magnitude of change and number of samples affected indicate the impact factor of a given feature.

Feature importance has enabled the identification of signal-specific features in \cite{SAYEDISMAIL20223539}. However, their approach utilised different features for ECG and PPG, as such, an intra-signal comparison could not be conducted, which would provide insights into the commonality of features between ECG and PPG, motivating the intra-signal feature importance analysis provided in this work.

Feature importance can also provide insights into the variance of features per affective state, valuable for the creation of tailored emotion-specific approaches. In \cite{filippini_chiara}, a neural network is used for classification, and the most important features were identified from the first layer's weights. These features were then evaluated to identify a statistical difference between affective states.

\subsection{ML Based Classification of Affective State}
A range of classifiers was selected to provide a holistic view of the classification performance using different architectures and the suitability of ECG and PPG for automated affective state detection.

Each classifier conducts a per-signal classification on each dataset, where 20\% of the data acts as a hold-out test set, which is unseen data used to evaluate the final classifier. To ensure generalisability, five-fold cross-validation is utilised, transforming the remaining 80\% of data into ``folds'', enabling a per-fold classification. Subsequently, comparing the per-fold and average performance across the five folds enables the identification of the most robust and performant classifier. Finally, the most performant classifier is trained on the entire training set and evaluated against the hold-out set to assess expected performance in real-life classifications.

\section{Results and Discussion}
\label{sec:results}

\begin{figure}[h]

    \centering
    \includegraphics[width=\textwidth]{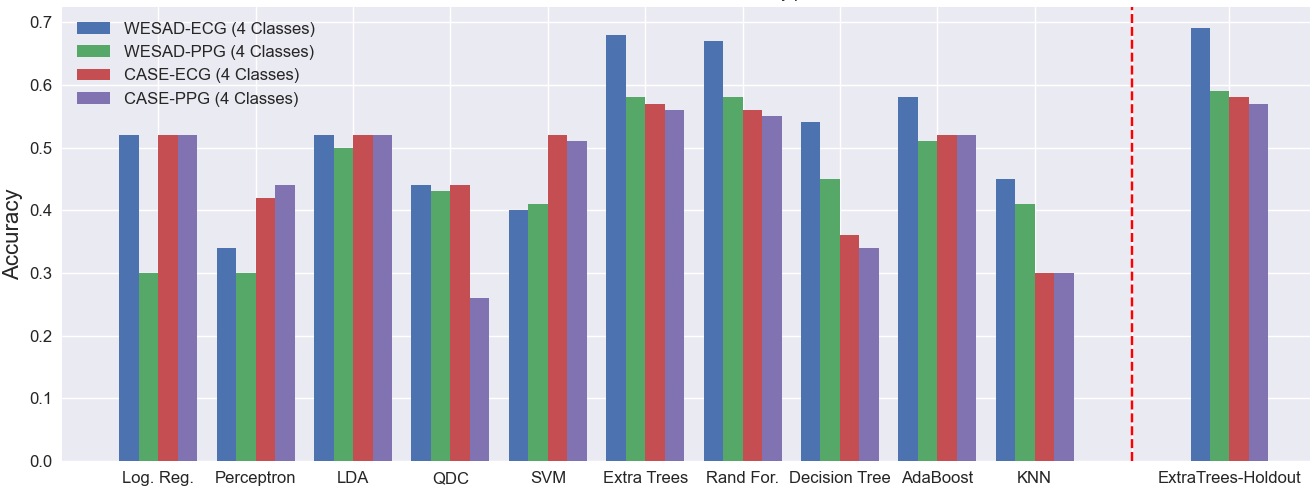}
    \caption{Mean cross-validation accuracy classifying affective states for model selection, and the performance on the holdout test set from the best performing classifier. The performance variance in ExtraTrees classifier between ECG and PPG reflects the degree of feature variance identified per dataset.}
    \label{fig:results_models}
    
\end{figure}

\subsection{HRV Feature Variance}

\begin{figure}[h]

    \centering
    \includegraphics[width=1\textwidth]{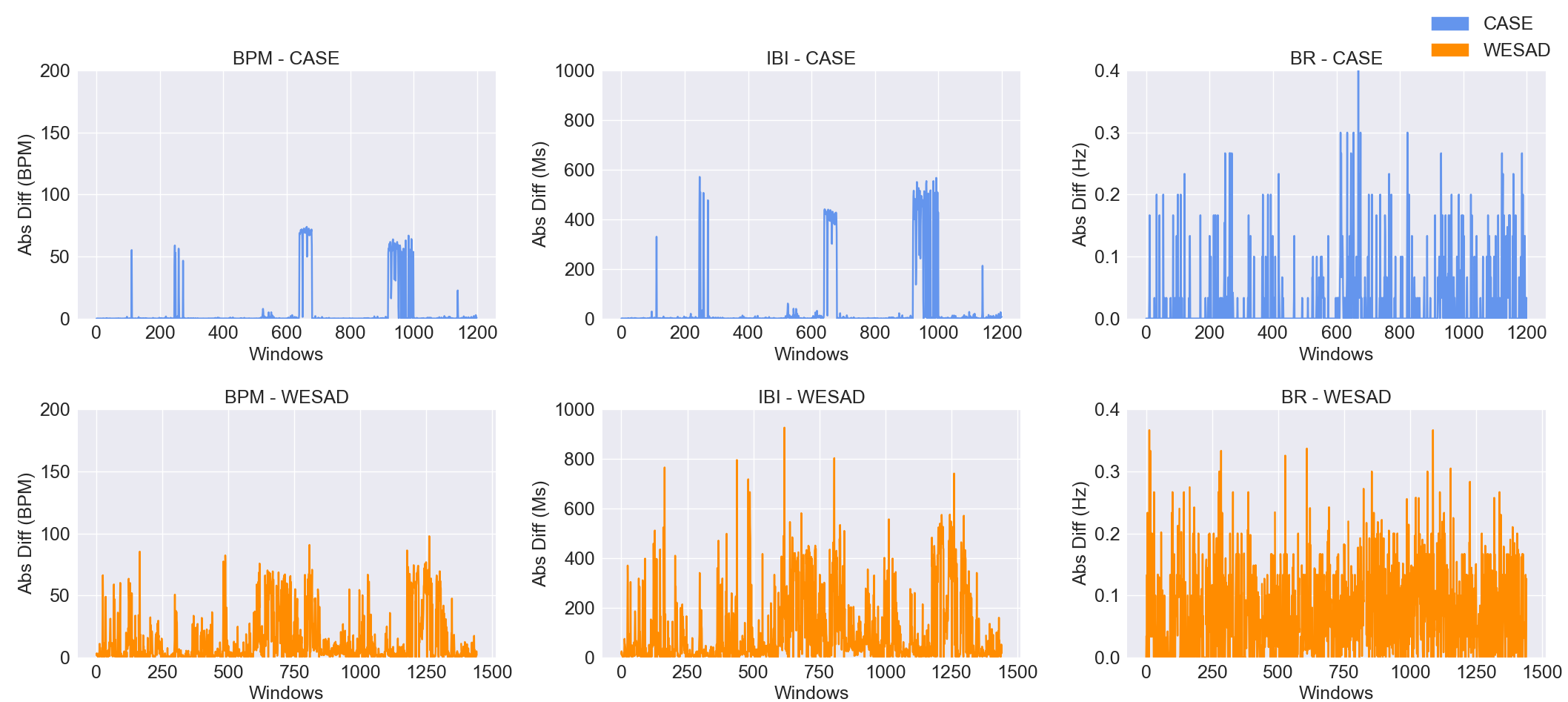}
    \caption{Absolute difference between ECG and PPG features: BPM, IBI and BR, for CASE and WESAD. High variance demonstrates unreliable feature computation in one of the signals due to signal noise or sensor differences.}
    \label{fig:stacked_feature_var}
    \vspace{-2em}

\end{figure}

\begin{figure}[h]
    \centering
    \subfloat[\centering Outliers Included]{{\includegraphics[width=0.47\textwidth]{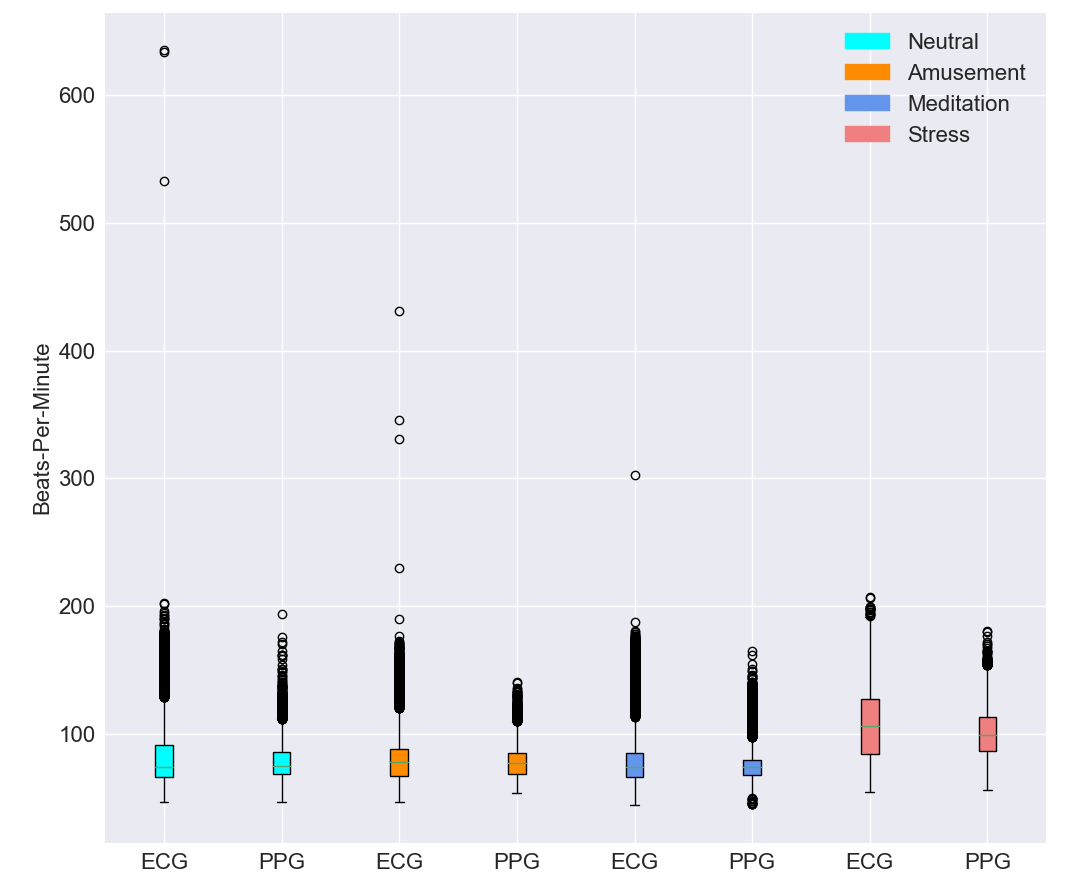}}}%
    \qquad
    \subfloat[\centering Outliers Excluded ]{{\includegraphics[width=0.47\textwidth]{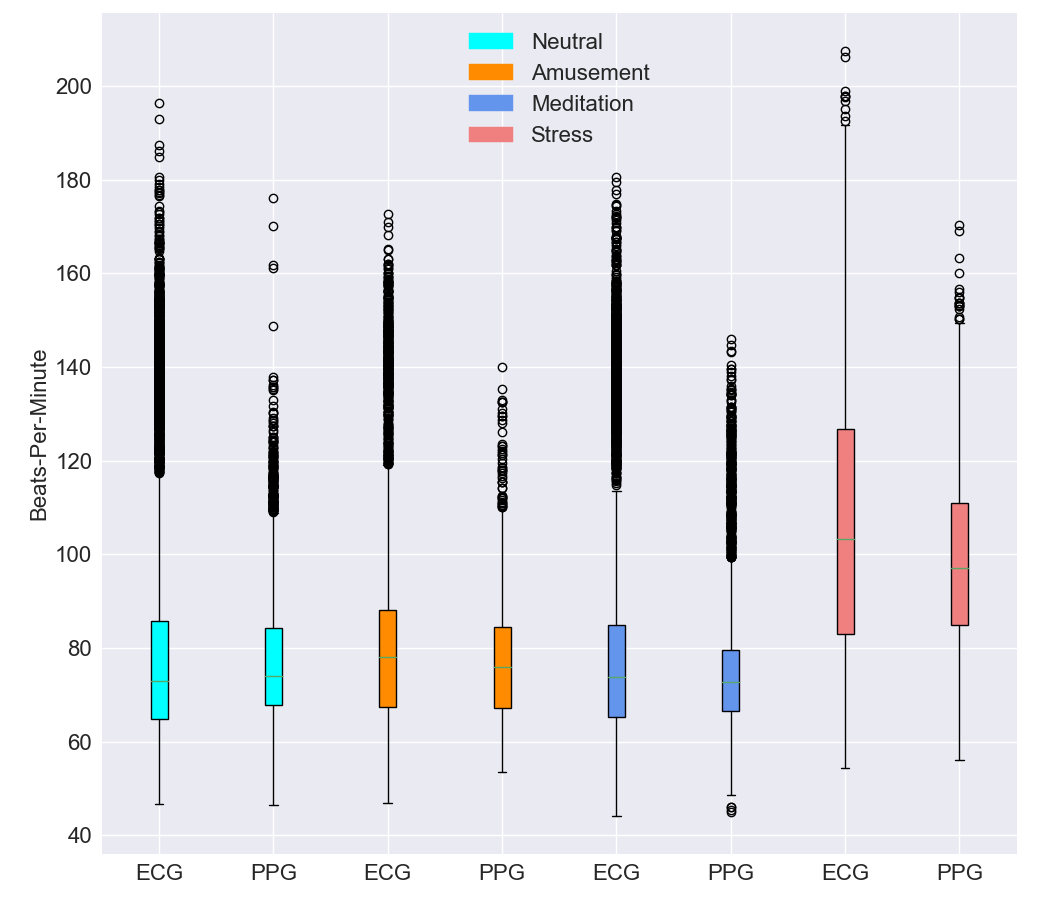}}}%
    \caption{Inter-signal and inter-affective state variance for BPM in WESAD, including and excluding outliers. Note in (a) the presence of outliers with a BPM of over 300 occurring in ECG indicating abnormal signal activity. Additionally, in (a, b), a visible overlap in neutral, amusement, and meditation occur, demonstrating the degree of similarity in these states.}%
    \label{fig:wesad_box_plots}%
    \vspace{-1em}
\end{figure}

The wearables' sample rate disparity (See Table \ref{tab:datasets}) is evident in the inter-signal feature variance results depicted in Figure \ref{fig:stacked_feature_var}. The reduced sample rates in WESAD result in slightly decreased granularity of ECG data and significantly in PPG data compared to CASE. A higher fluctuation in feature variance occurs in WESAD in terms of magnitude and frequency, stemming from the high sample rate disparity.

In CASE, beats per-minute (BPM) and inter-beat interval (IBI) contain a small variance with substantial spikes relative to the average. These variances occur in isolated data segments and are likely caused by electrode disconnection, movement, or subject-specific factors, visible in Figure \ref{fig:stacked_feature_var} at approximately window numbers 250, 550 and 900. Such occurrences may benefit from additional signal processing to reduce noise and improve feature computation accuracy.

Interestingly, the breathing rate (BR) feature exhibits a high deviation between signals in both datasets. This deviation indicates that at least one of the signals is unreliably computing BR, likely due to the wrist and finger placement of the PPG sensors.

A low degree of inter-affective state feature variance was identified between WESAD baseline, amusement, and meditation states for all features indicating these states are difficult to distinguish as depicted in Figure \ref{fig:wesad_box_plots}. Statistically similar features negatively impact automated classification, as the classifier struggles to differentiate between the classes. This impact is demonstrated by the reduced performance in multi-class classifications (58\%-69\%) as compared to the OVR performance depicted by the ROC curves (ROC Area: 0.70-0.95) in Figure \ref{fig:roc_curves}. This performance increase validates the utility of OVR classifications when classifying affective states that are difficult to differentiate due to statistical similarities.

\subsection{HRV Feature Importance}
\begin{figure}[tbp]
  \centering
  \begin{minipage}[b]{1.0\linewidth}
    \resizebox{0.40\textwidth}{!}{
    \begin{tabular}[b]{@{}ll@{}}
      \toprule
        Feature & Abbrv. \\\midrule
        Beats Per Minute & BPM \\
        Interbeat Interval & IBI \\
        Std dev. of RR Intervals & SDNN \\
        Std dev. of successive diff. & SDSD \\
        RMSE of successive diff. & RMSSD \\
        Proportion of diff. $<$ 20ms & pNN20\\
        Proportion of diff. $<$ 50ms & pNN50\\
        Median absolute dev. of RR intervals & MAD\\
        Estimated breathing rate & BR\\
        Poincare analysis & SD1, SD2,\\
        & S, SD1/SD2\\
        \midrule
        Dev: Deviation,
        Diff: Difference\\
      \bottomrule
    \end{tabular}
    }
    \qquad
    \centering
     \includegraphics[width=0.51\textwidth]{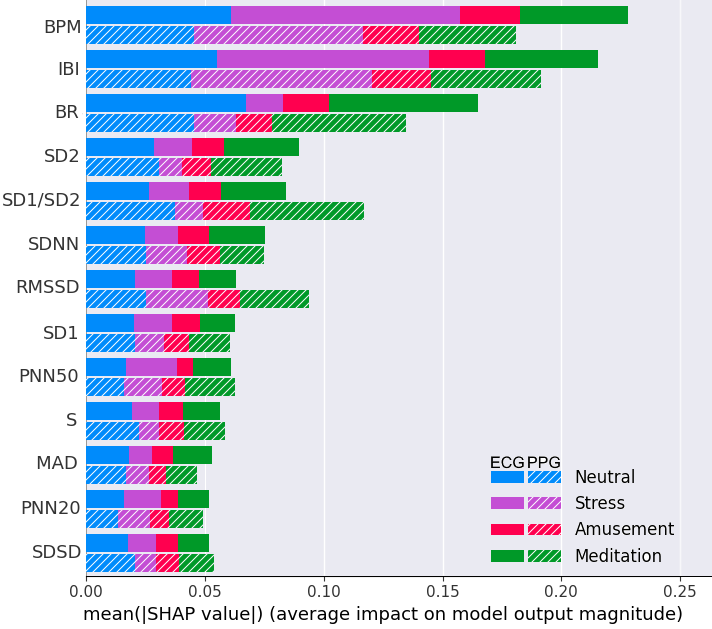}
  \end{minipage}
  \vspace{.5ex}
  \caption{\textbf{Sh}apley \textbf{A}dditive ex\textbf{P}lanations (SHAP) Feature Importance from ExtraTrees classifying WESAD signals.}
  \label{fig:feature_diff_WESAD}
\end{figure}

Analysing the SHAP values per feature indicates that BPM, IBI, and BR have the most significant impact on classification for both signals, as demonstrated in Figure \ref{fig:feature_diff_WESAD}. The remaining features exhibit inconsistent influence between the signals. Most notably, standard deviation 1 divided by standard deviation 2 (SD1/SD2) and room-mean-square of successive differences (RMSSD) exhibit higher impact in PPG as opposed to ECG.  This demonstrates the need for assessing feature importance on a per-signal basis to identify which features are most informative for use in tailored signal-specific classification approaches.

Certain features demonstrate varying impacts across affective states, indicating the presence of affect-specific features. For example, BPM and IBI exhibit high impacts on the class ``stress'', indicating their suitability for stress detection approaches. Assessing feature importance per-affective state provides an informative analysis of feature utility for affect-specific approaches. 

The high feature importance of BPM for ``stress'' is due to statistical distinction to the other affective states in the inter-affective state feature variance, as depicted in Figure \ref{fig:wesad_box_plots}. This demonstrates the benefit of assessing inter-affective state feature variance and feature importance to gain insights to aid the creation of affect-specific approaches.

\subsection{Automated Affective State Classification Variance}

Finally, the selected classifier is trained on the initial 80\% of data and classifies the hold-out set to assess expected performance in real-life classifications. The ExtraTrees classifier (ET) was selected as the most performant classifier from the model selection, where it was trained on 80\% of the training data and evaluated on the remaining 20\%. Notably, ET exhibits an increased performance when evaluated on the hold-out set as it was trained on all available training data. The full model comparison and ET hold-out performance is depicted in Figure \ref{fig:results_models}. Interestingly, the classifier performance variance between ECG and PPG is similar to the degree of the inter-signal feature variance identified per dataset.

\begin{figure}[!h]
    \centering
    
    \subfloat[\centering WESAD]{{\includegraphics[width=0.47\textwidth]{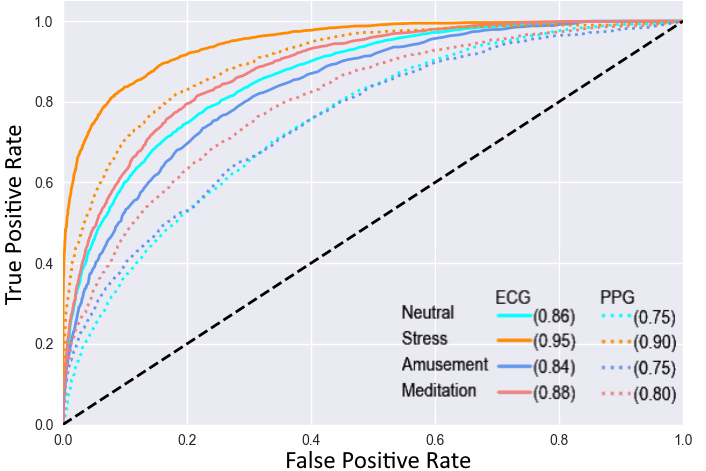}}}%
    \qquad
    \subfloat[\centering CASE ]{{\includegraphics[width=0.47\textwidth]{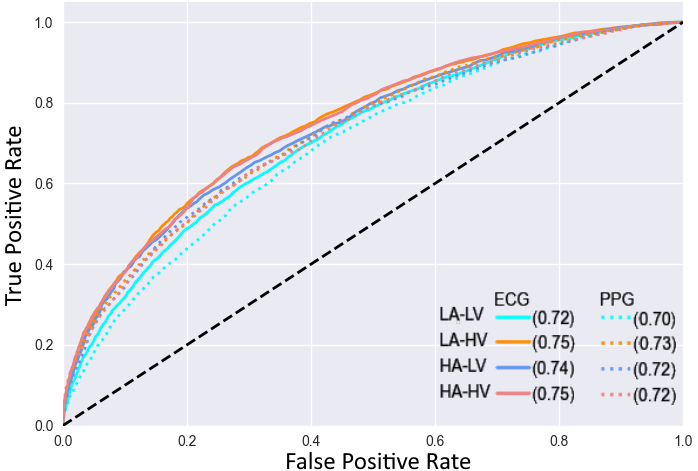}}}%
    \caption{ROC Curves from ExtraTrees representing the OVR classification variance between ECG and PPG}%
    \label{fig:roc_curves}%
    
\end{figure}

In contrast with the state-of-the-art \cite{sarkar_self-supervised_2021,Goshvarpour2020}, the performance achieved is lower for ECG and PPG; however, this work focuses on the analysis and understanding of variances between the signals for affective analysis rather than achieving high classification accuracy.
Analysing the ROC curves from ET demonstrates the true and false positive rates per signal for each affective state, see Figure \ref{fig:roc_curves}. On average, ECG demonstrates increased capabilities for affective classification by achieving a higher ROC area than PPG, varying with a range of 0.02-0.11. The increased performance via OVR demonstrates the benefit of identifying and overcoming the effects of similar affective states to achieve greater classification performance.

\section{Conclusions}
\label{sec:conclusions}
The inter-signal classification performance disparity mirrors the degree of feature variance between signals from both datasets. Specifically, WESAD exhibited a high feature variance, which explains the higher disparity in classification accuracy and ROC area per signal. Conversely, a lower inter-signal feature variance and a lower disparity in the performance measures occurred for CASE. This demonstrates the utility of inter-signal feature variance in identifying inconsistent computations of features stemming from sensor differences or abnormal signal activity, which negatively impact classification performance. These occurrences are likely to be more frequent in the ambulatory analysis due to motion artefacts and uncontrolled usage of wearables.

Furthermore, inter-affective state feature variance enables the identification of affective states that contain a similar distribution of features, which causes classification confusion. To counter this, the similar states are aggregated into an OVR classification problem, leading to increased performance, demonstrated by the ROC area per affective state.

Feature importance identifies BPM, IBI, and BR as the most impactful features for affective classification across ECG and PPG. Notably, the remaining features exhibit inconsistent impacts, specifically SD1/SD2 and RMSSD, which demonstrate a greater impact in PPG, warranting the exploration of signal-specific features. Analysing statistical measures to understand the inter-affective state feature variance indicates that certain features provide a greater degree of affect-specific information beneficial for tailored applications.

This work contributes an empirical analysis of data variances in ECG and PPG acquired using wearables and the impact on affective state classification. Therefore, enabling practitioners to make informed decisions when creating ML pipelines for affective state classification. The code-base will be made open access on Github (\url{https://github.com/ZacDair/Emo-Phys-Eval}), enabling automated feature variance analysis from each of these perspectives in a combined manner, regardless of data acquisition methods. While this approach analyses handcrafted features, it can also be utilised with automatically extracted features.

Future work will expand the analysis by utilising additional datasets to provide greater insights into the variances stemming from data collection devices, affective states, and population differences. In addition, an extended analysis will be conducted using additional features and methods to further inform the development of ML pipelines for affective state detection.
%
%
%
%

\bibliographystyle{splncs04}
\bibliography{ref}

\end{document}